\documentclass[10pt, amsmath, onecolumn, showpacs, article]{revtex4-2}

\usepackage{graphicx}
\usepackage{amsmath,amsfonts,amssymb,color}

\usepackage{hyperref}
\usepackage{footnote}
\usepackage{graphicx}
\usepackage{dcolumn}
\usepackage{bm}
\usepackage[caption=false]{subfig}

\begin{document}

\title[Using World Scientific's Review Volume Document Style]{Mixed Order Phase Transitions\label{ra_ch1}}

\author{David Mukamel}



\address{Department of Physics of Complex Systems, Weizmann Institute of Science\\
Rehovor 7610001, Israel}

\begin{abstract}
Mixed order phase transitions are transitions which have common  features with both first order and second order transitions. I review some results obtained in the context  of one of the prototypical models of mixed order transitions, the one-dimensional Ising model with long-range coupling that decays as truncated inverse square distance between spins. The correspondence between this model and the Poland Scheraga model of DNA denaturation, a subject to which Michael Fisher made substantial contributions, is then outlined. 
\end{abstract}

\maketitle


\section{Joining Michael Fisher as a postdoc at Cornell}\label{forward}
I was a postdoctoral fellow of Michael Fisher at Cornell in the years 1975-1977. Before arriving in Cornell I was a research fellow at Brookhaven National Lab where I applied symmetry considerations of Landau theory  to argue that $n$-vector spin models with $n\ge 4$ are in fact realized in a large variety of physical systems. Renormalization group analysis of these models thus provides their classification into universality classes. Furthermore, it suggests that when an accessible stable fixed point is lacking they should display a first order transition. 

At the time Michael was  skeptical about the use of Landau theory, beyond it being perhaps a useful but uncontrolled approximation. So upon my arrival in Cornell he asked me to present my work in one of his famous Tuesday seminars. The seminar turned into a tutorial series of three lengthy and grueling lectures where Michael kept posing questions and making infinitely many comments and suggestions.
After the third seminar he asked me to his office for a chat in which at some point he said, you know Landau theory is under appreciated in the West, and then he added candidly "...and maybe I am to blame". 
We ended up collaborating on a number of studies combining Landau's symmetry considerations with renormalization group calculations together with my old friend and his then student Eytan Domany. We have, for example, suggested a physical realization of the three state Potts model \footnote{For an impressive piece of art generated by Michael for this project where he displays his artistic drawing skills, see PRL 37, 565 (1976) and in the contribution of Eytan Domany in this volume, arXiv:2301:11086.} and demonstrated how a symmetry breaking field can turn a first order transition into a continuous one.

Cornell at the time was an exciting and intelectually stimulating place to be, and working with Michael was a unique experience.
I have learned a lot from Michael's non-compromising approach to science, how to try to select and pursue meaningful scientific problems and how to conduct research. My stay with Michael has deeply influenced my research throughout my career, and I am grateful for that.

\section{Introduction}\label{ra_sec1}

In the present article I briefly review some results on mixed order phase transitions, a subject to which Michael has made important contributions. Mixed order  phase transitions, also known as first order critical transitions or hybrid transitions, particularly in the percolation community,  are phase transitions which have common features with both first and second order phase transitions. In particular, they display a discontinuity of the order parameter as is common in first order phase transitions, and a diverging correlation length as is typical in second order phase transitions. They have been observed theoretically in a variety of models and found experimentally in some physical systems. Examples include the denaturation transition of DNA \cite{wartell1985, poland1966, fisher1966, Kafri2000,kafri2002},  jamming transitions \cite{gross1985,toninelli2006,jeng2007, toninelli2007,schwarz2006}, $k$-core percolation \cite{chalupa1979,dorogovtsev,park2020,lee2017universal}, non-enclave percolation \cite{sheinman2015}, models of wetting \cite{fisher1984,blossey1995},
networks dynamics \cite{bassler2015}, models of moving condensates \cite{whitehouse2014}, the one dimensional Ising model with long range interactions that decay as $1/r^2$ with the distance between the interacting spins \cite{thouless1969,dyson1971,anderson1969,anderson1970,anderson1971,fisher1982,cardy1981,aizenman1988} and a truncated interaction modification of this model \cite{bar2014PRL,bar2014JSAT,barma2019}.

While in some of the models the correlation length $\xi$ was found to diverge exponentially rapidly with the deviation, $\epsilon$, from criticality as $ \xi \sim e^{\epsilon^{-\mu}}$, in other models the divergence follows a power law, $\epsilon^{-\nu}$, with $\mu$ and $\nu$ some positive exponents. The former class includes models of jamming transitions \cite{toninelli2006} and the $1d$ Ising model with $1/r^2$ interaction \cite{anderson1971,cardy1981}, while the latter includes models of DNA denaturation \cite{poland1966,fisher1966,Kafri2000,kafri2002}, percolation transitions \cite{sheinman2015,dorogovtsev} and a truncated version of the $1d$ Ising model with long range interactions \cite{bar2014PRL,bar2014JSAT,barma2019}.

One of the prototypical models which exhibits a mixed order transition is the $1d$ Ising model with long-range interactions which decay as $\sim 1/r^2$ at large distance $r$. The model, later termed the inverse distance square Ising model (IDSI), was introduced by Thouless \cite{thouless1969} who argued that it possesses a phase transition with a jump in the magnetization at the transition. This magnetization discontinuity became known as the ``Thouless effect".  A closely related hierarchical model was  then introduced and solved by Dyson \cite{dyson1971} who demonstrated that magnetization discontinuity indeed takes place. An exact proof of the discontinuous nature of the transition in the IDSI model was eventually given by Aizenman et. al. \cite{aizenman1988}. In addition, scaling arguments \cite{anderson1971} and renormalization group analysis \cite{cardy1981} have shown that the correlation length $\xi$ diverges at the transition with essential singularity, $\xi \sim e^{b/\sqrt{T-T_c}}$, where $b$ is a constant and $T_c$ is the critical temperature. Although a lot is known about the thermodynamic behavior of this model, it is not exactly solvable. In $d=2$ and higher dimensions the model with $1/r^2$ interaction results in energy which scales super linearly with the volume. When properly scaled with the volume the model exhibits an ordinary continuous transition.

More recently, a modified version of the IDSI model was introduced \cite{bar2014PRL,bar2014JSAT,barma2019}. This model is exactly solvable and was shown to display a mixed order transition which is closely related to that of IDSI although with some differences. In what follows, this model is introduced and its main propertied are reviewed. I will then comment on the relation of this model to the Poland Scheraga model of DNA denaturation. I will also highlight the contributions of Michael to the understanding of the nature of mixed order transitions in general and to DNA denaturation in particular.



\section{The truncated inverse distance square Ising (TIDSI) model }

The TIDSI model is a truncated version of the IDSI model. Here, as in IDSI, the long range interaction between spins at distance $r$ apart decay as $1/r^2$ with the distance. However the interaction is confined to spins which lie within the same domain of either all up or all down spins. The Hamiltonian of the model is
\begin{equation}
	\mathcal{H}=-\Sigma_{i<j} J(i-j)\sigma_i \sigma_j I(i \sim j)-J_{NN} \Sigma_{i} \sigma_i\sigma_{i+1},
	\label{Hamiltonian}
\end{equation}
where $\sigma_i=\pm 1$,   $I(i \sim j)=1$ as long as sites $i$ and $j$ are in the same domain of either all up or all down spins and  $I(i \sim j)=0$ otherwise. The long range coupling is taken to be 
\begin{equation}
	J(r) \approx \frac{C}{r^2} \quad\, \text{for} \, \, r\gg1 ,
	\label{eq:J(r)}
\end{equation}
and, for simplicity, one may take periodic boundary conditions $\sigma_i=\sigma_{i+L}$ in a lattice of length $L$. Note that the indicator function $I(i \sim j)$ may be expressed in terms of the spin variables as 
\begin{equation}
	I(i \sim j)=\prod^{j-1}_{k=i} \delta_{\sigma_k \sigma_{k+1}}=
	\prod^{j-1}_{k=i} \frac{1+\sigma_k \sigma_{k+1}}{2},
\end{equation}
where $\delta_{\sigma_k \sigma_{k+1}}$ is the Kronecker delta function.
The Hamiltonian (\ref{Hamiltonian}) is thus not of the usual binary spin interaction form but rather it involves multi-spin interactions.

\begin{figure}[hb]
	\centerline{\includegraphics[width=11 cm]{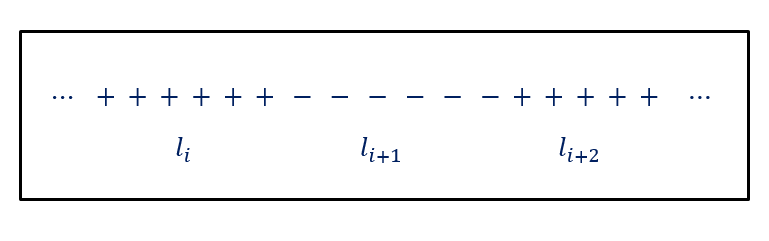}}
	\caption{ Microscopic configuration of the Ising model with the domain lengths indicated.}
	\label{fig: Isingconfig}
\end{figure}

It is convenient to express the Hamiltonian in terms of the domain length 
representation, where a domain is defined as a stretch of successive 
parallel spins (see Fig.~\ref{fig: Isingconfig}).  The long-range 
interaction in the first term operates only between pairs of spins which 
belong to the same domain, while the nearest neighbor interaction in the 
second term results in an energy cost for each domain wall. A typical 
configuration $\mathcal{C}$ is thus described by a set of domains with 
lengths $\{l_1,\, l_2,\, \cdots,\, l_N \}$ where the number of domains $N$ can 
vary from one configuration to another. In terms of these variables the Hamiltonian can be expressed as 
\begin{equation}       
	\mathcal{H} = \sum^N_{n=1} \mathcal{H}_n
\end{equation}
where
\begin{equation}       
	\mathcal{H}_n = -J_{NN}(l_n-2) -  \sum^{l_n}_{r=1}(l_n-r)J(r) \, .
\end{equation}
The variables $l_n$ are constrained by the system size $L$ to satisfy
\begin{equation}        
	\sum^N_{n=1} l_n = L \, .
\end{equation}
Using the form of $J(r)$ in Eq. (\ref{eq:J(r)}), one can estimate
the sum by replacing it by an integral as
\begin{eqnarray}
	\sum J(r) &\approx & a_0 - \frac{C}{l_n} \nonumber \\
	\sum rJ(r)& \approx & b_0 + C \, \ln l_n \, ,
	\label{sum.1}
\end{eqnarray}
where we have assumed that $l_n$ is large and kept the two leading order
terms for large $l_n$. This 
is justified
since we are interested in phenomena close to the critical 
line where domains are typically large.  
Dropping an overall unimportant constant,
one obtains the effective Hamiltonian 
\begin{equation} 
	\mathcal{H} = C \sum_n \ln l_n + \Delta N
\end{equation}
where the constant $C$ is the amplitude of the long-range interaction and 
$\Delta = 2J_{NN}+C + b_0$ acts as a chemical potential for 
the number of domains. 
The resulting Hamiltonian is that of a Coulomb gas of alternating sign charges in $1d$, where the interaction is confined to nearest neighbor charges only, and it is logarithmic in the distance, as is the case of charges in  $2d$. In this representation the charges correspond to the domain boundaries, and the domain lengths, $l_n$, are the distance between nearest neighbor charges.

The free energy and the resulting thermodynamic properties of the model have been calculated in recent studies and the mixed order nature of the transition has been clarified \cite{bar2014PRL,bar2014JSAT,barma2019}. In addition, a discussion of this model within the context of a different class of models exhibiting fluctuation-dominated phase ordering has been given by Barma {\it et. al.} \cite{barma2019} and in the contribution of Mustansir Barma in the present volume. Let us briefly outline this analysis.  The probability of a configuration $\mathcal{C}$, defined by the domain lengths $(l_1, \dots, l_N)$ and their number $N$ is given by its Boltzmann weight 
\begin{equation}        
	P(l_1,\, l_2,\, \cdots,\, l_N,\, N|L) = 
	\frac{y^N}{Z_y(L)}\, \prod^N_{n=1}\, 
	\frac{1}{l_n^c} \, \delta_{\sum^N_{n=1}l_n, \, L }  
	\label{jointd.1}
\end{equation}
where 
\begin{equation}
	c=\beta C \,\,\, \mbox{ ,} \,\,\, y=e^{-\beta \Delta} \, ,
	\label{c}
\end{equation}
and $\delta_{i,j}$ is the Kronecker
delta function that enforces the sum rule. The normalization constant
$Z_y(L)$ is the partition function given by 
\begin{equation}        
	Z_y(L) =  \sum^\infty_{N=1} y^N \sum^\infty_{l_1=1} 
	\ldots \sum_{l_N=1}^{\infty} \prod^N_{n=1} 
	\frac{1}{l_n^c}\delta_{\sum^N_{n=1} l_n, \,L} \, .
	\label{pf.1}
\end{equation}
Here we have ignored a geometrical factor which arises from the number of ways of placing the sequence of domains on the lattice. This factor results in a subleading contribution to the partition sum and hence it does not affect the thermodynamic properties of the model.
\begin{figure}[hb]
	\centerline{\includegraphics[width=11 cm]{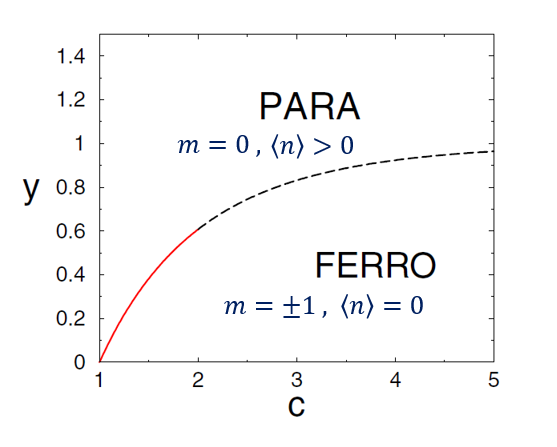}}
	\caption{ The $(c,y)$ phase diagram of the TIDSI model. The red solid line represents a continuous transition with $\langle n \rangle$ vanishing continuously at the transition. The dashed line represents a mixed order transition with a discontinuous order parameter $\langle n \rangle$ and diverging length. Note that the $m$ order parameter is discontinuous at the transition for any $c>1$.}
	\label{fig: phasediagram}
\end{figure}

In order to analyze the behavior of the model and particularly the phase diagram and the probability distribution of the domains length, we proceed by considering the grand canonical ensemble. The grand partition sum, $Q_y(z)$ is given by
\begin{equation}        
	Q_y(z) =  \sum^{\infty}_{L=1} z^L Z_y(L) \, ,
	\label{pfgf.1}
\end{equation}
where $z$ is the fugacity. Carrying out the sums in  Eqs. (\ref{pf.1}, \ref{pfgf.1}) one obtains
\begin{equation}        
	Q_y(z) =  \frac{y \phi_c(z)}{1- 
		y \phi_c(z)}
	\label{pfgf.2}
\end{equation}
with
\begin{equation}        
	\phi_c(z) =  \sum^{\infty}_{l=1} \frac{z^l}{l^c} \equiv {\rm 
		Li}_c\left(z\right)\, ,
	\label{phis.1}
\end{equation}
where ${\rm Li}_c(z)$ is the polylogarithmic function.
The sum in Eq.(\ref{phis.1}) converges for $z<1$, diverges for $z>1$ and it exhibits a singularity at $z=1$.

The phase diagram and the nature of the phase transition can be deduced by direct inspection of $Q_y(z)$. The partition sum of the model, $Z_y(L)$ is determined by the pole of $Q_y(z)$ when it exists.  The pole is obtained at $z^*$ which satisfies
\begin{equation}
	y\phi_c(z^*)=1 \, ,
	\label{z*}
\end{equation}
and it exists as long as $z^*<1$. In this case the partition sum is given by
\begin{equation}
	Z_y(L) \propto z^{*-L} \, .
	\label{Z_y}
\end{equation}
A phase transition takes place in the model at $z^*=1$, namely at $y$ and $c$ satisfying
\begin{equation}
	y\phi_c(1)=1 \, .
	\label{transition}
\end{equation}

The nature of the phases and the phase transition are simply found by calculating the distribution function , $P(l)$, of a single domain. This is done by summing over the lengths of all other domains in (\ref{jointd.1}) yielding
\begin{equation}
	P(l)=\frac{Z_y(L-l)}{Z_y(L)}\frac{y}{l^c} \,  ,
	\label{singled1}
\end{equation}
which, given (\ref{Z_y}), results in
\begin{equation}
	P(l)=\frac{yz^{*l}}{l^c} \propto \frac{e^{-l/\xi}}{l^c} \,,
	\label{singled2}
\end{equation}
as long as $z^*<1$. Here $\xi \equiv -1/{ \ln z^*}$  is a length scale which controls the typical domain size. For finite $\xi$ (namely for $z^*<1$), the equilibrium domains are of finite typical length controlled by $\xi$, and their average number $N$ scales with the system's length $L$. As a result, the domain's density $n=N/L$, is finite.
As $z^* \rightarrow 1$ the correlation length $\xi$ diverges and the density of domains (or charges) vanishes. One can think of two order parameters involved in this transition: the magnetization $m$ and the average number of domains $\langle n \rangle$. The two order parameters yield somewhat different results as to the nature of the phase transition, as is outlined below.

In the spin representation, a typical configuration in the paramagnetic phase is composed of a finite density of domains of alternating up and down spins, so that the overall magnetization vanishes. At the transition point, $z^*=1$, the domain's density vanishes, leading to a phase which is composed of a single domain of length $L$ with some small domains whose total length is subleading in $L$. This is a ferromagnetic phase  with average magnetization per spin $m=\pm 1$ for any $c>1$. The transition from the paramagnetic to this ferromagnetic phase was thus termed "extreme Thouless effect". The fact that the discontinuous order parameter is accompanied by a diverging correlation length, makes the transition of mixed order nature for any $c>1$.


\begin{figure}[hb]
	\centerline{\includegraphics[width=11 cm]{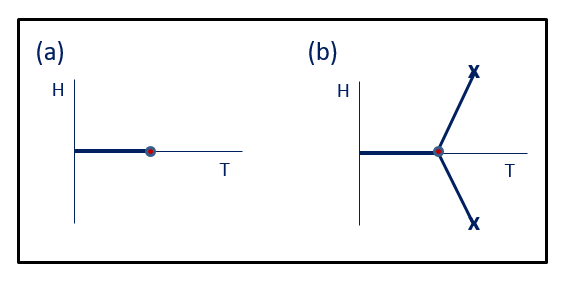}}
	\caption{Typical schematic $(T,H)$ phase diagram for (a) second order transition at $H=0$ and (b) first order transition at $H=0$ with three first order lines intersecting at a triple point. The two fork branches terminate at critical points marked x in the figure. Scaling arguments suggest that the phase diagram of the IDSI model is of the form of (a).  }
	\label{fig: HTIsing}
\end{figure}

\begin{figure}[hb]
	\centerline{\includegraphics[width=11 cm]{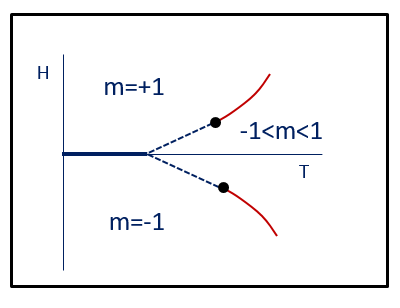}}
	\caption{Schematic $(T,H)$ phase diagram of the TIDSI model with $c>2$. Dashed lines represent lines of mixed order phase transition of the TIDSI type with $H$ dependent parameter $c(H)>2$, while red thin lines represent continuous transitions with $1<c(H)<2$. On the thin lines the magnetization $m$ approaches $m=\pm 1$ continuously while on the dashed lines it exhibits a discontinuous jump. The thick solid line at $H=0$ is a first order transition line on which the magnetization changes discontinuously from $+1$ to $-1$. }
	\label{fig: HTmixed}
\end{figure}

In order to explore the nature of the phase transition using the $\langle n \rangle$
order parameter let us consider the domain size distribution at the transition, $P(l)\propto 1/l^c$, which exists as a normalizable distribution for $c>1$. For $1< c \le 2$ the average domain length, $\langle l \rangle$, diverges, and thus the the average density of domains, $\langle n \rangle$ vanishes. This  indicates that  the order parameter of the transition, $\langle n \rangle$, goes continuously to zero at the transition. The transition is thus a usual second order transition with diverging correlation length and continuously vanishing order parameter. On the other hand for $c > 2$, $\langle n \rangle$ is finite at the transition, indicating that the transition into the ordered phase, at which $\langle n \rangle$ vanishes, involves a discontinuous change of the order parameter. Thus for $c > 2$ the model is of a mixed order nature, with a diverging length $\xi$ and a discontinuous order parameter $\langle n \rangle$. Note that the parameter $c$ of the model, given in (\ref{c}), is temperature dependent, and the nature of the transition is determined by the value of $c$ at the transition temperature, namely $c=\beta_c C$. The phase diagram of the model in the $y-c$ plane is given in Fig. \ref{fig: phasediagram}.

The critical behavior of the model can be extracted from the singularity of the polylogarithmic function at $z=1$.  In the disordered phase, where the domains density $\langle n \rangle$ is non vanishing and $z<1$, the fugacity is related to temperature variable $y$ ($\equiv \beta \Delta$) via $y\phi_c(z)=1$. Close to the transition point, and for $1<c<2$ the polylogarithmic function satisfies $\phi_c(1-\delta z)-\phi_c(1) \propto \delta z ^{c-1}$. Thus close to the transition, where $y=y_{crit} +\delta y$, one has $\delta y \propto \delta z^{c-1}$. The correlation length
$\xi=-1/{\ln (\delta z)}$ thus satisfies $\xi \propto \delta z^{-1}$, or $\xi \propto \delta y^{-\frac{1}{c-1}}$. 
Since $y$ is basically the temperature variable , $\delta y$ is proportional to the reduced temperature $t=(T-T_c)/T_c$ resulting in 
\begin{equation}
	\xi \propto t^{-\nu} \quad {\rm with} \quad \nu=1/(c-1) \, . 
	\label{nu}
\end{equation}
For $c>2$ the singular term in $\phi_c (z)$ is sub leading with $\phi_c (1-\delta z)-\phi_c(1) \propto z$ which yields $\nu=1$.  

The order parameter of the transition is $\langle n \rangle = 1/{\langle l \rangle}$. Using the domain length distribution (\ref{singled2}) one finds that for $1<c<2$ 
\begin{equation}
	\langle n \rangle = 1/\xi^{2-c} \propto t^{\frac{2-c}{c-1}} \, .
	\label{ n order parameter}
\end{equation}
For $c>2$ , the average domain length $\langle l \rangle $ is finite and the order parameter is discontinuous at the transition.

As mentioned above, in the IDSI model the correlation length diverges at the transition with an essential singularity rather than a power law. An insight into this behavior is gained by considering the model within the framework of the charge configurations. Here, as in the TIDSI model, a microscopic configuration is expressed as a sequence of alternating charges marking the domain boundaries. However, unlike TSIDI, the logarithmic interaction between the charges is not restricted to nearest neighbor charges only but is rather all to all. This makes the transition of a Kosterlitz-Thouless type, with its characteristic essential singularity. This has first been studied by Anderson {\it et al.} \cite{anderson1971} using scaling arguments and later elaborated by more detailed renormalization group approach by Cardy \cite{cardy1981}.

An interesting question is what is the effect of an ordering field $H$ on a mixed order phase transition. In ordinary second order phase transition an ordering field  removes the transition, as, for example is the case for the Ising model. On the other hand first order transitions are retained in the presence of an ordering field, resulting in a fork shape phase diagram in the ($T,H$) plane where three lines of first order transitions meet at a triple point (see Fig. \ref{fig: HTIsing}). In the case of the IDSI model it was first conjectured by Anderson {\it et al.} \cite{anderson1971} that the ($T,H$) phase diagram of this mixed order transition  is like that of an ordinary second order transition (see Fig. \ref{fig: HTIsing}a). Fisher, who has always been intrigued by first order transitions, developed together with Berker a scaling approach to first order transitions and argued that this is indeed the case and termed the transition ``first order critical point" \cite{fisher1982}. The phase diagram of the TIDSI model is quite different. In Figure \ref{fig: HTmixed} a schematic phase diagram of a model with $c>2$ is illustrated. In this figure the $H=0$ mixed order transition survives the ordering field. The phase diagram has a fork shape with two lines of mixed order phase transitions meeting the $H=0$ first order transition line. The mixed order lines are characterized by a continuously varying parameter $c(H)$. This parameter decreases with $H$ until it reaches the value $c=2$ where the transition becomes continuous. On the mixed order sections of the branches the magnetization changes discontinuously from some value in the paramagnetic phase to $|m|=1$ in the ferromagnetic phase. However on the continuous section of the branches the magnetization changes continuously to its $|m|=1$ value in the ferromagnetic phase (for details see \cite{bar2014JSAT}).

\begin{figure}[hb]
	\centerline{\includegraphics[width=11 cm]{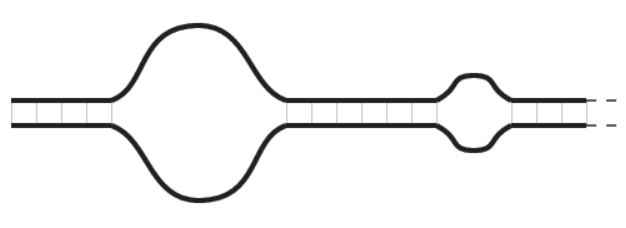}}
	\caption{ Schematic configuration of DNA molecule with alternating bound segments and open loops.}
	\label{fig: DNAconfig}
\end{figure}

\section{Comment on DNA denaturation}

DNA denaturation corresponds to melting of double stranded DNA, a phase transition at which the two bound strands of the molecule become separated. This transition has been extensively studied experimentally and theoretically for over sixty years by now. The canonical model used to study the transition is the Poland-Scheraga (PS) model introduced in the 1960's \cite{poland1966}. Within this model a DNA molecule is schematically viewed at an alternating sequence of segments of bound strands and denatured loops of unbound single strands (see Fig. \ref{fig: DNAconfig}).  The statistical weight of a microscopic configuration, which is characterized by the number of segments and length of each one of them is determined by the following simplifying assumptions: each bound pair carries a (negative) binding energy $u$ so that the statistical weight of a bound segment of length $l$ is $\omega=e^{-\beta u l}$. For simplicity all bound pairs carry the same energy and the bound segment is non-flexible so that it carries no configurational entropy. A segment of length $l$ of unbound pairs constitutes a loop of length $2l$. The loop carries no energy, however it is viewed as composed of a flexible string and thus it possesses configurational entropy which has to be taken into account is calculating the statistical weight of the DNA configuration. In the PS model as introduced by Poland and Scheraga, loops are considered as fully flexible and they are modeled  as non-interacting random loops. In the large loop limit the statistical weight of a loop of length $l$ in $d$ dimensions is
\begin{equation}
	\Omega(l) \propto \frac{s^l}{l^c} \, \, ,
	\label{omega}
\end{equation}
where $s$ is some $d$ dependent constant and $c=d/2$. Here $1/l^c$ represents to probability of a random walk in dimension $d$ to return to the starting point after $l$ steps.

Equation (\ref{omega}) implies that as in the TIDSI model, the statistical weight of a loop may be expressed by two opposite charges interacting logarithmically as $c\ln l$, leading to the same type of transition as in TIDSI. In particular, the loop length distribution takes the form of Eq. \ref{singled2}, leading to no transition for $c \le 1$, continuous transition for $1< c <2$ and a mixed order transition for $c \ge 2$. A major distinction between the two models is that while the parameter $c$ in TIDSI is temperature dependent, and thus nonuniversal, in the PS model $c$ is universal. It is independent of temperature and of any type of short range interactions that may exist between the two strands. The prediction of the PS model for DNA denaturation is that $c=3/2$ in $d=3$, implying a continuous transition. This result disagrees with the experimental observations which suggest that the transition is discontinuous. 

An important step in resolving this discrepancy was made by Michael Fisher in 1966 \cite{fisher1966}, who observed that exclusion interaction between the strands in a loop, which was neglected in the PS analysis, should be taken into account. This interaction is long ranged and can thus affect the value of $c$. Noting that the end-to-end distance $R$ of a self avoiding random walk of length $l$ scales as $R \sim l^\nu$, the probability of the walk to return to the origin becomes $1/l^{d \nu}$, making $c=d\nu$.  Given that in $d=3$ the Flory exponent is estimated to be $\nu \approx  0.583 $, the parameter $c$ increases from $1.5$ in the noninteracting model to $\approx 1.75$ when exclusion is taken into account. While the predicted transition is still continuous, the inclusion of the exclusion interaction is a step in the right direction. 

This approach was later taken a step forward by Kafri {\it et al.}\cite{Kafri2000,kafri2002} who noted that the exclusion interaction between a loop and the rest of the chain modifies $c$ even further. Modeling the rest of the chain as a linear flexible polymer one can apply a renormalization group approach developed for evaluating the entropy of a general polymer network \cite{duplantier1986,duplantier1989,schafer1992} to study the length distribution of a self avoiding loop embedded in a chain. It was found that the loop size distribution is of the same form as that of an isolated loop but with a modified parameter $c$ which takes the value  $c\approx 2.11$ in $d=3$ dimensions. This makes the denaturation transition of mixed order type with a discontinuous order parameter, in line with what experiments suggest.

\section{Summary}

Mixed order phase transitions have been suggested to exist in a variety of theoretical models of physical systems. In this paper some properties of this intriguing class of phase transitions are reviewed and  their derivation is outlined in the framework of the TIDSI model. This model is exactly solvable and thus provides useful insight into the mechanism leading to a transition with discontinuous order parameter together with a diverging correlation length. The model can be expressed in terms of charges of alternating sign in $1d$ whose interaction is restricted to nearest neighbor only. This feature is found to lead to a power law divergence of the correlation length. When the restriction is removed and the interaction becomes all-to-all, the model represents the IDSI model, which displays a Kosterlitz-Thouless type mixed order transition with exponential divergence of the correlation length. It would be of great interest to explore the relevance of this picture of interacting charges to models in higher dimensions, such as models of jamming transitions.

\section*{Acknowledgments}
I thank Amir Bar, Mustansir Barma, Yariv Kafri, Satya N Majumdar, Luca Peliti and Gregory Schehr with whom I have collaborated over the years on studies of properties and peculiarities of mixed order transitions. I also thank Eytan Domany and Martin Evans for comments on the manuscript.
This work was supported by a research grant from the Center for Scientific Excellence at the Weizmann Institute of Science.

\bibliographystyle{unsrt}
\bibliography{ws-rv-sample}

\end{document}